\documentclass{aa}    
\usepackage{graphicx}
\usepackage{txfonts}
\begin{document}
\title{Re-condensation from an ADAF into an inner disk - 
the intermediate state of black hole accretion?}
\author{F. Meyer \inst{1}, B.F. Liu \inst{2} and E. Meyer-Hofmeister
\inst{1}}
\offprints{Emmi Meyer-Hofmeister; emm@mpa-garching.mpg.de}
\institute{Max-Planck-Institut f\"ur Astrophysik, Karl-
           Schwarzschildstr.~1, D-85740 Garching, Germany\\
           \email{frm@mpa-garching.mpg.de;emm@mpa-garching.mpg.de}
           \and
           National Astronomical Observatories/Yunnan Observatory,
           Chinese Academy of Sciences, P.O. Box
           110, Kunming 650011, China\\
           \email{bfliu@ynao.ac.cn}
           }

\date{Received: / Accepted:}

\abstract
 {Accretion onto galactic and supermassive black holes occurs in
 different modes, which are documented in hard and soft spectral
 states, commonly attributed to an advection-dominated flow (ADAF)
 inside a truncated disk and standard disk accretion, respectively. 
 At the times of spectral transition an intermediate state is
 observed, for which the accretion flow pattern is still unclear.}
 {We analyze the geometry of the accretion flow when the mass flow
 rate in the disk decreases (soft/hard transition) and evaporation of gas into 
 the coronal flow leads to disk truncation.}
 {We evaluate the physics of an advection-dominated flow 
 affected by thermal conduction to a cool accretion disk underneath.}
 {We find re-condensation of gas from the ADAF into the underlying inner
 disk at distances from the black hole and at rates, that depend on the 
 properties of the hot ADAF and vary with the mass accretion
 rate. This sustains an inner disk for longer than a viscous decay
 time after the spectral transition occurred, in accordance with the 
 spectra that indicate cool gas in the neighborhood of the accreting 
 black hole. The model allows us to understand why Cyg X-1 
 does not show hysteresis in the spectral state transition luminosity
 that is commonly observed for X-ray transient sources.} 
 {Our results shed new light on the complex mass flow pattern during
 spectral state transition.}

 
\titlerunning {Re-condensation from an ADAF into an inner disk}
\maketitle


\section{Introduction}
Accretion onto compact objects occurs in different modes.
With the investigation of radiatively inefficient
accretion flows (Narayan \& Yi 1994), the observed spectra of low-mass
X-ray binaries and Active Galactic Nuclei (AGN) could be understood, at
low luminosities, as arising from a hot advection-dominated accretion flow
(ADAF). The ADAF model or its variants provide qualitative
understanding of this hard spectrum from optically thin emission 
(Narayan 2005). At high luminosities the accreting gas forms
a standard accretion disk (Shakura \& Sunyaev 1973) reaching inward to 
the last stable orbit, producing a soft multi-color black body
spectrum.

But from the large number of observations for X-ray binaries 
during the last years (McClintock \& Remillard 2006) it became 
obvious that for many sources the spectra document a highly complex 
appearance of accretion processes. Intermediate states were found, 
especially in connection with the change between the hard and soft spectral
state (in both directions), or even the very high state, in both kind
of sources, neutron star and black hole transients (Psaltis
2006). Hardness-intensity diagrams, e.g. the diagram for the
outburst of GX 339-4 in 2002/2003 (Belloni et al. 2005),
show that the evolution of hardness
with count rate is different during outburst rise and decay,
indicating differences in the accretion flows via disk and ADAF.
Similar changes in hardness are visible in diagrams for XTE
J1650-500 (Homan et al. 2003) and 4U 1630-47,
(Tomsick et al. 2005). The variety in diagrams for the hard/soft
change can be seen from the recent compilation of Gierli\'nski \&
Newton (2006). 
Kalemci et al. (2003) in their work on black hole X-ray transients
during outburst decay characterize the observed features as
``several global patterns of evolution for spectral and temporal 
parameters before, during, and after the transition''. 
In general the Fe K$\alpha$ line observed from stellar black holes and
AGN, the reflection component, and also the timing properties might point to 
emission from cool gas in the innermost region or from the edge of the disk.

Our work concerns the pattern of accretion
flow at the time of spectral state transition. It is not yet
clear how the accretion flow changes from one mode to the other, 
e.g. from disk accretion inward to the last stable orbit to a
truncated disk with a hot advection-dominated flow in the inner
region. We investigate this change starting from the picture that
gas evaporated that from the disk feeds the coronal flow. What happens 
when the mass flow rate in the disk decreases (as during decline from an
outburst) and becomes lower than the evaporation rate maximum? Then
only a hot flow exists. We discuss that the disk reaching inward to
the last stable orbit breaks up where the evaporation efficiency is
maximal (at a distance of some hundred Schwarzschild radii
($R_{\rm{S}}=2GM/c^2$)). Farther inside, the cool disk would still
exist. however,But not being fed by a connection to the outside disk, such a 
relic inner disk would disappear in a viscous time. Apparently observed
intermediate states last longer. An inner disk can then only survive if 
it is kept up by condensation of gas from the ADAF into the disk.
Since these intermediate states are poorly understood at the
present time we consider, as a first step, a new situation: Can cool
gas exist underneath an ADAF? We discuss the physics for this kind
of accretion flow in an analytical model. To include more detailed
physics, an evaluation by numerical simulation has to be performed. We
cannot yet answer in detail the question of how such cool gas could be 
recognized in the observed spectra. 

In the past the difficulty in
interpreting the spectral observations of transient low-mass X-ray binaries
(LMXB) led to the question of whether the observations indicate ``two
independent simultaneous accretion flows'', as discussed by Smith et
al. (2002) and Pottschmidt et al. (2006). But whatever the
geometrical configuration is in which accretion would happen 
simultaneously via a hot flow and via a cool disk, at the same  
distance these flows would not be independent of each other. This
especially concerns thermal conduction. An interaction with cold
clumps embedded in the hot flow causing thermal conduction of the ADAF
in the radial direction was discussed by Yuan \& Zdziarski (2005).

One of the key questions in the study of the hot flow above cool gas is
whether heat can be drained from the upper ADAF and be radiated away, 
so that matter condenses onto the cool disk. We investigate the
physical properties of an ADAF affected by thermal conduction to an 
accretion disk below the ADAF. In particular, the energy balance of the
hot flow at low height above the cool disk surface where ion and
electron temperature couple is important. (For a discussion of
the problem see Liu et al. 2006.)

A peculiar aspect of such evaporation and condensation processes
is the wide disparity between the amount of mass contained in the ADAF
and in the disk, at each moment at each distance. This is a direct
consequence of widely disparate temperatures in the two media. The
velocities of the inward flow, proportional to the temperature, are
thus also different by orders of magnitude and, for comparable
mass flow rates lead to the very different column densities in ADAF
and disk. How then can such a tenuous ADAF or corona, by condensation
or evaporation affect and support a much more massive disk
underneath? Even as a column of the ADAF loses only a fraction of its
small mass during the short time that it passes over the disk surface,
the underlying disk receives the contributions of many such columns before
it has moved significantly inward, and the accumulated mass
received can amount to a significant fraction of its own mass content.

In Sect.2 we discuss the diversity of the outburst light curves of
X-ray transients and the observations for an intermediate 
spectral state. In Sect.3 we describe the accretion flow pattern 
related to the soft/hard spectral change. The physics of the hot flow 
affected by thermal conduction is treated in Sects.4 to 6. 
Using an analytical procedure we show in Sect.7 how the energy balance 
in the radiating layer
close to the disk surface determines whether evaporation or
condensation occurs. In Sect. 8 we determine the  
re-condensation rate of gas from the ADAF onto an inner relic disk. 
The model allows us to evaluate how the condensation depends on the mass
flow rate in the ADAF and the parameters involved. We draw
conclusions regardingunder which circumstances condensation
would lead to an inner weak disk below the ADAF, appearing in the
spectrum as soft contribution. In Sect. 9 we show how our model can 
explain why Cyg X-1 does not have hysteresis in the
transition luminosity, which is observed for all X-ray transient sources. In
Sect. 10 we compare our model with observations for an intermediate state of
X-ray transients, discuss the inner edge of the disk and the
application to disks in AGN. Conclusions are given in Sect. 11.

A better understanding of the pattern of the accretion flow should also  
give insight to the observed complex timing characteristics, the 
occurrence and type of quasi-periodic oscillations (QPO) and the radio
connections to the spectral states. 
Jet production corresponding to special accretion modes (Gallo \& 
Fender 2005) also is part of the picture, arising from the accretion physics.

\section{The outburst light curves of X-ray transient sources}
\subsection{The cause of the mass flow changes}
McClintock \& Remillard (2006)
examined 18 confirmed black hole binaries. In many sources the
accretion rate seems to increase and decrease several times until
finally after an outburst a quiescent phase follows. The large
diversity of light curve features was 
documented by Chen et al. (1997), only a few
sources showing the simple pattern of  ``fast rise -
exponential decay'' (FRED), but in many sources 
the luminosity varies in a complex way (see also
 the discussion by Gierli\'nski and Newton 2006).
 Two issues together 
create the variety of changes of luminosity and spectral states:
(1) The outburst probably is caused by the ionization instability, 
as the dwarf nova outbursts (Meyer \& Meyer-Hofmeister 1984). 
The mass flow in the disk is related to
the spread of heating and cooling waves initiating the changes between
hot and cool disk structure. A larger disk size in X-ray transients 
together with irradiation leads to a more complex pattern in mass flow and
outburst light curve (for the effect of a larger disk see accretion
disks in symbiotic stars, Duschl 1986);  (2) The mass flow rate in the 
disk resulting from this
outburst behavior is the key parameter that determines the accretion
mode, for high rates via an accretion disk everywhere (soft spectrum), 
for lower rates a truncated outer disk and an ADAF inside (hard
spectrum). 

\subsection{The intermediate spectral state}
McClintock \& Remillard (2006) redefined the X-ray states of black hole 
X-ray binaries in terms of quantitative criteria that utilize both
X-ray energy spectra and power density spectra. They took as intermediate
the spectra that did not belong to low/hard or high/soft state, nor
to the very high state (sometimes a large part of the observation,
e.g. 50\% throughout the 2002-2004
outburst of 4U 1630-47, Tomsick et al. 2005). 
We do not consider the intermediate state in
the change to the very high/soft state. This latter state is of different
physical nature (see the discussion by Fender et al. 2004, and the
model of disk fragmentation by Meyer 2004).

The observations show that generally the intermediate state phase
during the hard/soft transition (in outburst rise) is shorter than
that during the soft/hard transition (in outburst decay). For the
former transition durations, of 3 to 11 days were found (Corbel et
al. 2004, G\"o\u{g}\"u\c{s} et al. 2004, Belloni et al. 2005).

Our work focuses on the accretion flow pattern during the latter transition.
Kalemci et al. (2003) analyze the evolution of spectral and temporal
properties of several galactic black hole transients during outburst
decay:  XTE J1650-500, GRO J1655-40, XTE J1748-288, XTE J1755-324, 4U
1630-47, XTE J1550-564, XTE J1859+226 and GX 339-4 (for this
source see also Revnivtsev et al. 2000) and the spectral analysis shows
a decrease of the disk component until it becomes unobservable with 
PCA /RXTE within 15 days after the transition. Similar times were
found for GRS 1758-258 (Smith et al. 2001, Pottschmidt et al. 2006). 
Cyg X-1 is a special case with an intermediate state lasing for weeks
(Cui et al. 1997, Zdziarski et al. 2002). 
The results for neutron star sources show the same trend with shorter 
timescales (Barret \& Olive 2002).

\section{The accretion flow pattern at the 
soft/hard spectral transition: a gap in the disk?}
In earlier work on accretion onto compact objects (Meyer
\& Meyer-Hofmeister 1994, Meyer et al. 2000, Liu et al. 2002, see also
R\'o\.za\'nska \& Czerny 2000a, 2000b) we found that 
above the cool accretion disk a hot corona always is built up, fed by matter
evaporating from the disk. The efficiency of evaporation, a siphon
flow, depends on the distance from the compact star. The evaporation rate
increases with decreasing distance to the center until a maximum 
$(\dot M_{\rm{evap}})_{\rm{max}}$ is reached at $R_{\rm{evmax}}$. 
The mass flow rate in the disk determines whether the coronal flow
is important. If it is higher than $(\dot M_{\rm{evap}})_{\rm{max}}$ 
the disk is essentially unaffected and reaches
inward to the last stable orbit, and the hot flow is unimportant (this is
the usual situation when the luminosity is high during an outburst,
see also discussion in Sect. 8.3). 
If the mass flow rate in the disk is low,
the disk is truncated where the evaporation has transfered all gas to the
hot flow/ADAF (the usual situation in quiescence). According to the mass flow
rate, spectral transitions occur. Note that these two transitions occur at 
different luminosities, an observational feature known as hysteresis in
X-ray transient light curves. This is theoretically understood as due
to the different hard/soft character of the irradiation from the
central region in the two cases (Meyer-Hofmeister et al. 2005, Liu et al. 2005). 

\begin{figure}
\centering
\includegraphics[width=8.3cm]{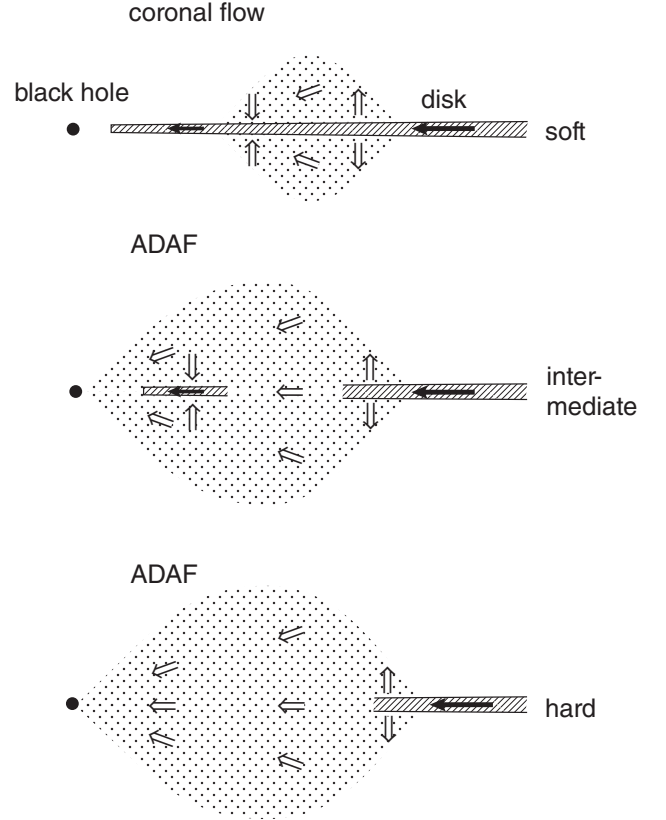}
\caption[.]{\label{coronalflow}
Accretion flow in soft, intermediate and hard spectral
state (with decreasing mass flow rate). Upper panel: In the soft state
accretion occurs via the disk, some gas
is evaporated to a coronal flow, but re-condenses onto the disk; in
the inner region only the disk flow exists, which causes the soft spectrum. 
Middle panel: In the intermediate state the ADAF and an inner disk 
contribute to the spectrum. Lower panel: In the hard state only the
hot ADAF contributes to the spectrum.
} 
\end{figure} 

When during the decrease from an outburst the mass flow rate in the
disk becomes equal to the evaporation rate maximum, the disk truncation
starts at the distance $R_{\rm{evmax}}$, and a gap appears, filled by an
ADAF, as illustrated in Fig. \ref{coronalflow}
. With the further decrease of mass flow the gap should widen due to 
evaporation. What happens to the left-over inner cool disk? 
Diffusion causes a spread inwards and outwards, decreasing
the amount of disk matter due to accretion onto the compact object and
evaporation in outer regions. Only if matter can condense from the
ADAF onto the disk can such a disk survive longer than the viscous
time for these processes, which is only a few days in the inner disk. 
If the disks still exists the spectrum can show contributions
from the ADAF and a reflection from the left-over inner cool disk.

\section{ADAF properties}
As our analysis of the hot ADAF above a left-over cool disk will show,
the upper part, i.e. the main part in vertical extension, is not
affected by thermal conduction to the cool disk below. We make use of the
solutions of self-similar advection-dominated flows by
Narayan \& Yi (1995). The properties of these flows scale with 
black hole mass, mass flow rate and distance from
the black hole. They further depend on the viscosity and
the assumed magnetic field strength. Following Narayan et al. (1998)
the magnetic pressure is written    
\begin{equation}\label{p_mag}
p_{\rm m}=(1-\beta)\rho{c_{\rm s}}^2
\end{equation}
with $\beta$ ratio of gas pressure
to total pressure, $\rho$ density and $c_{\rm s}$
isothermal sound speed. Shearing box simulations of
turbulence driven by the magneto-rotational instability in a
collisionless plasma by Sharma et al. (2006) yield $\beta$ values
around 0.8. We use this value in our analysis. 
As the ratio of specific heats of
the magnetized plasma we take $\gamma=(8-3\beta)/(6-3\beta)$ (Esin
1997) though the true value would require a more detailed analysis.

For the chemical abundance a hydrogen mass fraction of 0.75 was used.
The solutions for pressure, electron number density, viscous dissipation of
energy per unit volume $q^+$ and isothermal sound speed are 
(Narayan \& Yi 1995) 
\begin{eqnarray}\label{scaled}
p&=&1.87\times 10^{16}\alpha^{-1}m^{-1}\dot m r^{-5/2} 
\rm{g cm^{-1} s^{-2}} \nonumber ,\\
n_e& =&5.91\times10^{19}\alpha^{-1}m^{-1}\dot m r^{-3/2} \rm{cm^{-3}} \\
q^+&=& 2.24\times10^{20}m^{-2}\dot m r^{-4}\rm{ergs cm^{-3} s^{-1}}
\nonumber,\\
c_{\rm s}^2&=&1.67\times 10^{20} r^{-1} \rm{cm^2 \,s^{-2}} \nonumber.
\end{eqnarray} 
where $\alpha$ is the viscous coefficient,
$m$ the black hole mass in units of
solar mass $M_ \odot$, $\dot m$ the mass flow rate in units of the 
Eddington accretion rate $\dot M_{\rm Edd}=1.39\times 10^{18}m\,{\rm
g/s}$, and $r$ the radius in units of the Schwarzschild radius 
$R_{\rm S}=2.95\times 10^5 m
\,\rm{cm}$.
The ion number density is $n_i=n_e/1.077$.
In an ADAF, ion and electron temperatures $T_i$ and $T_e$
closely follow  
\begin{equation}
T_i+1.077 T_e =1.98\times10^{12} r^{-1}\rm K 
\end{equation}
and, if $T_e$ is much smaller than $T_i$ this value can be taken for
$T_i$ alone. 
Besides viscous heating, $q^+$ compressive heating
$q^c=\frac{1}{(1-\beta)}q^+$ (Esin 1997) is important.

\section{The electron thermal profile}
In a pure ADAF, at radii $r<10^3$, the coupling between ions and electrons
becomes poor. Almost all of the accretion energy stays with the ions, which
reach near virial temperatures, the flow has a high vertical
extent, and the density is low (see review by Narayan et al. 1998) . The
electrons cooled by bremsstrahlung and synchrotron(-Compton) radiation
stay at temperatures around $10^{9.5}K$. Radiation losses are very weak.

\begin{figure}
\centering
\includegraphics[width=7.cm]{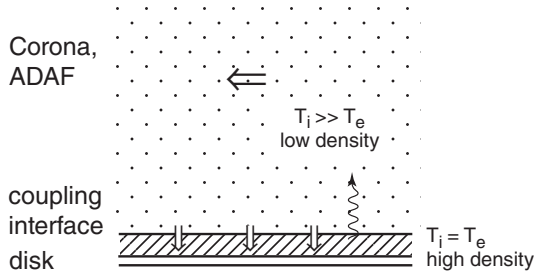}
\caption[.]{\label{disk+ADAF}
The hot two-temperature ADAF together with
the radiating one-temperature layer below the coupling interface.
}
\end{figure}

This situation changes if there is a disk below the
ADAF. Due to the large temperature difference between the hot ADAF and
the cool disk, thermal conduction results in further cooling of the
electrons. While the ion temperature is not much affected, the electron
temperature drops with height until, near the bottom, a temperature
$T_e=T_{\rm cpl}$ is reached at which coupling between ions and electrons
becomes efficient, and from then on ion and electron temperatures are the
same, $T_i=T_e=T$, as illustrated in Fig. \ref{disk+ADAF}
.

The electron temperature has practically no influence on the dynamics
of the flow and the thermodynamics of the ion gas and is
subrelativistic, $kT_e <m_ec^2$ ($k$ Boltzmann constant, $m_e$
electron mass, $c$ speed of light). The rate of energy
transfer from ions to electrons is given by Stepney (1983). Since in
the two-temperature advection-dominated hot flow the ions are at a
much higher temperature than the electrons a simplified
formula (Liu et al. 2002) can be used
\begin{eqnarray}\label {qie}
q_{ie} & = & 3.59\times 10^{-32} {\rm{g cm^{5} s^{-3} deg^{-1}}}
\, n_e n_i T_i
{\left(\frac{k T_e}{m_e c^2}\right)}^{-\frac{3}{2}}
   \nonumber \\
        & = & 1.05\times 10^{35} {\rm{ g cm^{-1}s^{-3}deg^{3/2}}}
T_e^{-3/2} \alpha^{-2}m^{-2} \dot m^{2}r^{-4}.
\end{eqnarray}.

Coupling of ions and electrons is reached when viscous and compressive
heating of the ions is balanced by the collisional heat transfer from
the ions to the electrons, $q_{ie}=q^+ + q^c$.  This then
determines  the coupling temperature $T_{\rm {cpl}}$. Since the
coupling occurs at low height the density there is higher than
the height averaged value (Eq. ({\ref{scaled})), by a factor
$2/{\sqrt \pi}$ (vertical density distribution
$n=n_0\,\rm{exp}(-{z^2}/{H^2})$). This yields the coupling temperature as
\begin{equation}\label{T_cpl}
T_{\rm {cpl}}=1.98\times 10^9 \alpha^{-4/3}\dot m^{2/3}= 1.24\times 10^9 K,
\end{equation}
the numerical value for $\alpha$=0.2 and $\dot m$=0.02 (see
Sect. 8.1).

In the case considered here, conductive cooling limits the peak electron
temperature in the ADAF to a value $T_m$ somewhat lower than the limit
where radiation losses balance collisional heating. The
synchrotron radiation is optically thick, which makes the synchrotron
radiation losses drop steeply with temperature (Narayan \& Yi 1995,
Mahadevan 1997). As bremsstrahlung also decreases with
temperature we may neglect to a first approximation the radiation losses
and calculate the resulting thermal electron profile from the balance
between thermal conduction and collisional heating alone,
\begin{eqnarray}\label{heatflux}
F_{\rm c}&=&-\kappa_0 T_e^{5/2}dT_e/dz,\\
\frac{dF_{\rm c}}{dz}&=&-q_{ie}(T_e).
\end{eqnarray}
For the thermal conductivity coefficient $\kappa _0$
we take the standard value (for the effect of $\kappa _0$
on the coronal structure see Meyer \& Meyer-Hofmeister 2006).
Formally multiplying the left sides and the right sides of these two
equations by each other and multiplying by $dz$ reduces this to a
first order differential equation for $F_c(T_e)$ with the solution
\begin{equation}\label{Fc2}
F_{\rm c}^2=\kappa_0 (K n_i n_e T_i)
(T_{\rm m}^2-{T_e}^2)
\end{equation}
with $K= 1.64\times 10^{-17} \rm {g cm^{5}s^{-3}deg^{1/2}}$.
The integration constant $T_m$ is the maximal electron temperature
reached with height. Note that the downward directed heat flow
has a negative value.  Integration of $dz/dT_e$ (Eq.{\ref{heatflux}})
using  Eq.(\ref{Fc2}) yields the electron temperature profile
in the ADAF,
\begin{equation}\label{z(Te)}
z_{\rm m}-z=\sqrt{\frac{\kappa_0}{K n_i n_e T_i}}{T_{\rm m}}^{3/2}
\int^1_{T_e/T_{\rm m}}\frac{x^{5/2}}{\sqrt{1-x^2}}\,dx.
\end{equation}.

For the low height $z=z_{\rm cpl}$ where $T_e$ has dropped
to the coupling temperature $T_{\rm cpl}$, in Eq.(\ref{z(Te)})
we may neglect $z_{\rm cpl}$ compared to $z_{\rm m}$, and with
$T_{\rm cpl}^2/T_{\rm m}^2$ small compared to one, the integral
may be extended from zero to one to give 0.719. Eq.(\ref{z(Te)}) then
yields the maximal electron temperature in the ADAF,
\begin{equation}\label{T0}
T_{\rm m}^{5/2} = 1.39\, z_{\rm m} \,{(K n_i n_e T_i/ \kappa_0)}^{1/2}.
\end{equation}

For the height $z_{\rm m}$ one may take the vertical
scale height for $n_e n_i$ which is $1/\sqrt{2}$ of the density
scale height $c_{\rm s}/\Omega_{\rm K}$ ($\Omega_{\rm K}$
Kepler angular velocity) of  Narayan et al. (1998),
\begin{equation}\label{z_m}
  z_{\rm m}= 1.27\times 10^5\,m\,r\ {\rm cm}.
\end{equation}
With this value we obtain $T_{\rm m}$ from  Eq.(\ref{T0}).
Eq.(\ref{Fc2}), neglecting $T_{\rm cpl}^2$ compared to $T_{\rm m}^2$,
finally yields the thermal flux drained from the ADAF that enters 
the coupling interface at $z=z_{\rm cpl}$ from above,
\begin{equation}\label{Fc}
F_{\rm c}^{\rm{ADAF}}=-
\kappa_0^{3/10}\, {(K n_i n_e T_i)}^{7/10}
{(1.39\, z_{\rm m})}^{2/5}.
\end{equation}

\section{The coupling interface}
In the case of no mass exchange between ADAF and disk,  Eq.(\ref{Fc}) gives
the heat flux that enters at $z=z_{\rm cpl}$ into the radiating
layer from the ADAF above. This is the borderline case
considered by Liu et al. (2006). However if mass condenses
from the ADAF onto the disk it carries the high thermal
heat content of the ADAF ions to the interface. Energy
conservation across the interface yields
\begin{equation}\label{F_cpl1}
F_{\rm{cpl}}+ \dot m_z \frac{\gamma}{\gamma-1}\frac{1}{\beta}
\left(\frac{\Re T}{\mu} \right)
=F_{\rm{c}}^{\rm{ADAF}}+ \dot m_z
\frac{\gamma}{\gamma-1}\frac{1}{\beta}\left(\frac{\Re T_i}
{\mu_i}\right)
\end{equation}
The factor $\frac{1}{\beta}$ arises when we include magnetic work and
internal energy advection together with the corresponding gas terms
in the simplified ADAF parameterization of the magnetic field.
Here $F_{\rm{cpl}}$ is the heat flow leaving the interface on the
lower side, $\Re$ is the gas constant, $T$ is the common temperature
of ions and electrons on that  side, and $\dot m_z$ is the mass flow
rate of the condensing gas per unit area. (It has the dimension of
${\rm g}{\rm cm}^{-2}{\rm s}^{-1}$ and should not be confused
with the non-dimensional ADAF mass accretion rate $\dot m$ and
similar quantities defined later.) In condensation flow it has a
negative value. The heat flow $F_{\rm{cpl}}$ entering the
radiating layer below the ADAF from above is thus increased over
$F_{\rm c}^{\rm{ADAF}}$,
\begin{equation}\label{F_cpl2}
F_{\rm{cpl}}=F_{\rm{c}}^{\rm{ADAF}}\left[1-\frac{\dot m_z}{\dot
m_z^*}\right]
\end{equation}
where
\begin{equation}\label{mz}
\dot m_z^*=\frac{\gamma-1}{\gamma}\beta  \frac{-F_{\rm{c}}^{\rm{ADAF}}}
{\frac{\Re T_i}{\mu_i}(1-\epsilon)}
\end{equation}
is a normalization mass flow rate per unit area (positive and of
the same physical dimension 
as $\dot m_z$). It depends on the ADAF
parameters. The ratio
\begin{equation}\label{epsilon}
\epsilon=\frac{\mu_i T_{\rm{cpl}}}{\mu T_i}
\end{equation}
is a small number, typically of the order of $10^{-2}$. An example for  
temperature and density distributions is shown in Fig. \ref{logzcpl}
.

(In the case of evaporation ($\dot m_z >0$) the situation
at the boundary between the radiating layer and the ADAF is
different. While for a descending flow, collisional coupling
immediately heats the electrons and adds to their thermal
flux, in an ascending flow, on passing through the boundary,
the ions becoming uncoupled must first be heated up by friction
to the ADAF temperature before they are fully incorporated
in the ADAF. Here we consider only the condensation case.)

\begin{figure}
\centering
\includegraphics[width=6.cm]{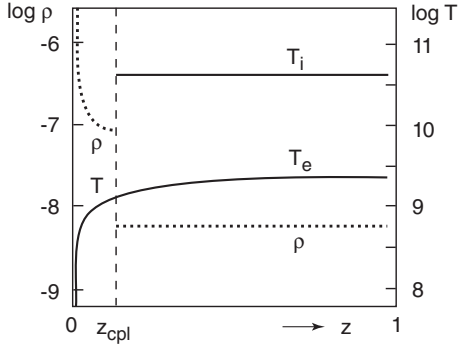}
\caption[.]{
\label{logzcpl}
Changes of ion temperature, electron temperature and
density at the distance of 80 Schwarzschild radii and for a mass flow
rate of $\dot m$=0.02 in the ADAF. The vertical height $z$ is measured
in units of the vertical scale height for $n_e n_i$.
}
\end{figure}

\section{The radiating layer below the ADAF}
The parameters of the ADAF determine what happens in the layer
between the coupling interface and the disk surface.

Case (1): If the pressure is high, the
density in the lower layer also is high. Then the
conductive flux drained from the ADAF is already efficiently radiated
away at some height before the disk surface is reached. Radiative
cooling however continues and must be served
   by additional heat released from gas
descending through the temperature profile to the disk surface, i.e.
matter condenses from the ADAF onto the disk.

Case (2): If the pressure is low, only part of the
conductive flux can be radiated away; the remaining part is taken up
by cool disk matter heating up as it rises through the temperature
profile, i.e. matter evaporates from the disk into the corona.

A borderline case in between, case (3), occurs if the ADAF pressure
allows the thermal heat flux drained from the ADAF to be
radiated away exactly on reaching the disk surface.
Then there is neither condensation nor evaporation, i.e. no mass
is exchanged between disk and ADAF.

\subsection{The energy balance in the radiating layer}
We derive the condition for the different
cases from the energy balance in the radiating layer (compare Liu et
al. 2002). We use a simplified form of the energy equation,
keeping only the dominant contribution of internal heat, pressure
work, and thermal conduction, together with bremsstrahlung cooling
$n_e n_i \Lambda(T)$,

\begin{equation}\label{energy}
\frac{d}{dz} \left[\dot m_z \frac{\gamma}{\gamma-1} \frac{\Re T}{\mu} +
F_c \right] = -n_e n_i \Lambda(T).
\end{equation}
We express density by temperature and (constant) gas pressure
$\beta\,p$. This value is taken at the bottom of the ADAF
region, and, as for the density, is slightly higher than the
vertical mean pressure (Eq.(\ref{scaled})). Assuming free-free
radiation for $T_e \ge 10^{7.5}\rm{K}$,  $n_e n_i \Lambda(T)$ becomes
$\frac{0.25}{k^2}{(\beta\,p)}^2 bT^{-3/2}$ with
$b=10^{-26.56} \rm{g\,cm^{5} s^{-3} deg^{-1/2}}$ (Sutherland \& Dopita
1993).
For simplicity we use this law also for smaller $T$. The
justification is that because of the very steep temperature profile
below $10^{7.5}$K such regions contribute only a negligible amount to
cooling of this layer (Liu et al. 1995).
Contributions of gravitational energy release, frictional heating, and
side-wise advection of mass and energy can be neglected for a small
extent of this layer. Likewise, kinetic energy is negligible since at
high density the flow is highly subsonic.

To solve the second-order differential equation Eq.(\ref{energy}),
we use $T$ as the independent variable and define a new
dependent variable
$g(T)\equiv \kappa_0T^{3/2}dT/dz=-F_{c}/T$. We now obtain the
first-order differential equation
\begin{eqnarray}\label{gT}
g \frac{dg}{d\ln T} &=& \frac{0.25\beta^2 p^2}{k^2} \kappa_0 b
   +  \dot m_z \frac{\gamma}{\gamma-1}
\frac{\Re}{\mu}g - g^2
\nonumber \\
&=& - (g-g_1)(g-g_2),
\end{eqnarray}
with
\begin{equation}\label{g_1}
g_1={\dot m_z\over 2} {\gamma\over \gamma-1} {\Re\over \mu}+
\sqrt{\left({\dot m_z\over 2} {\gamma\over \gamma-1} {\Re\over
\mu}\right)^2
+\frac{0.25\beta^2 p^2}{k^2} \kappa_0 b}
\end{equation}
and $g_2$ differing from $g_1$ only by the sign of the square root.

This equation has to be solved with the upper and lower boundary
conditions
$g=\frac{-F_{\rm{cpl}}}{T_{\rm{cpl}}}$ at $T=T_{\rm{cpl}}$
and $F_{\rm c}=0$ at $T=0$. In reality, the temperature does not drop
to zero at the disk surface but the solution is practically independent
of the exact values at the lower boundary (except for a very
narrow range in $z$  at the bottom) as long as flux and
temperature there become small compared to those at the
upper boundary, with no consequence for our results. $\dot m_z$ is the
Eigenvalue to be determined.

The only solution that fulfills the lower boundary condition is the
singular solution $g(T)=g_1=const$, i.e. a linear relation between
$-F_{c}$ and $T$. The ratio of these two quantities  at the upper
boundary determines the value of $g_1$. This requires

\begin{equation}\label{mdotz}
\dot m_z = \frac{\gamma-1}{\gamma}
\frac{-F_{\rm{c}}^{\rm{ADAF}}\left(1-\frac{\dot m_z}{\dot m_z^*}\right)} 
{ \frac{\Re T_{\rm {cpl}}}{\mu}}
\left[1- C \frac{1}{1-(\frac{\dot m_z}{\dot m_z^*})^2} \right]
\end{equation}

with

\begin{equation}\label{C}
C = \kappa{_0} b
\left(\frac{0.25\beta^2 p^2}{k^2}\right)
\left(\frac{T_{\rm {cpl}}}{F_c^{\rm{ADAF}}}\right)^2 \,.
\end{equation}

Eqs.(\ref{mdotz}), (\ref{mz}), and
(\ref{epsilon}) yield a quadratic equation for $\dot m_z$,
\begin{equation}\label {1mdotz}
\left(1-\frac{\dot m_z}{\dot m_z^*}\right)^2
-\epsilon \left(1-\frac{\dot m_z}{\dot m_z^*}\right) -(1-\epsilon)C
= 0\,,
\end{equation}
with the solution
\begin{equation}\label {mdotzmdotz}
\frac{\dot m_z}{\dot m_z^*}=1-
\frac{\epsilon}{2}-\sqrt{\frac{\epsilon2}{4} +(1-\epsilon)C}]\,.
\end{equation}
The quantity $C$ of Eq.(\ref{C}) compares the radiation loss
in the thermal profile with the heat flux drained from the ADAF.
For $C=1$ Eq.(\ref {mdotz}) gives $m_z=0$, the borderline
case discussed by Liu et al. (2006). For $C<1$, the radiation
losses are too weak, $\dot m_z$ is positive and the heat flow
is used up to heat the evaporating gas to the (un-)coupling
boundary temperature $T_{\rm cpl}$. If, on the other hand,
$C>1$, the gas in this layer is efficiently cooled, sinks
down and condenses.

\section{The mass exchange between ADAF and disk}
\subsection{The borderline case: no mass exchange, $C=1$}
The value of $C$ depends on the ADAF parameters $\alpha$,
$\beta$, $\dot m$ and $r$, but not on $m$. This shows that
the same situation holds for supermassive and
stellar black holes alike. Also note a significant dependence
on $\beta$, $C$ scales almost with $\beta ^2$ (Eq.(\ref{C})).
For $\beta$=0.8 we get
\begin{equation}\label{ccrit}
C=0.96\, \alpha^{-28/15}  \dot m ^{8/15} r^{-1/5}.
\end{equation}
The radius $r=r_c$ at which $C=1$ marks the transition
between outside evaporation and inside condensation of
an inner disk below the ADAF,
\begin{equation}\label{rc}
r_c= 0.815\,\alpha^{-28/3} \dot m ^{8/3}.
\end{equation}
Here, we take $\alpha$=0.2 (corresponding to $\alpha$=0.3 in
Shakura-Sunyaev notation) but note the extremely
strong dependence of $r_c$ on the choice of this parameter.
Since $\alpha$ incorporates several underlying physical
processes of unresolved detail its exact
value is not known. Various  values of the above order of
magnitude are quoted in the literature (see Narayan et al.1998).
The re-condensation process discussed here occurs in
connection with spectral state transitions for which
Maccarone (2003) found $\dot m=0.02$ from a detailed
investigation of these transitions in X-ray binaries. For these
values $C$ becomes 1 at a distance of about 80 Schwarzschild
radii, an order of magnitude discussed for inner disks
in intermediate states (Ibragimov et al. 2005). The dependence
of  $C$ on mass flow rate and distance obtained  allows us to derive
conditions for the existence of an inner disk below the ADAF.

\subsection{Condensation in the inner region.}
It is an interesting question as to how the fraction of the ADAF that
re-condenses onto the inner disk and then flows inward via
the disk depends on the ADAF parameters. For this we add
all contributions from the distance $R_c=r_cR_{\rm S}$ where
condensation begins down to an inner cut-off radius $R_i=r_iR_{\rm S}$,
e.g. the radius of the innermost stable orbit around a non-rotating
black hole, $r_i=3$. As discussed later, the inner
edge may well lie at some larger distance. Like the ADAF accretion
rate, the total condensation rate will be measured in Eddington
accretion rate. (In our definition condensation rates have a negative
sign.) Using Eq.(\ref{mdotz}), (for $\epsilon \ll 1$ )  the total rate
of condensation onto both sides of the disk becomes
\begin{eqnarray}\label{condrate}
\mid \dot m_{\rm{cond}}\mid &=&\int^{R_{\rm{c}}}_{R_{\rm{i}}}
\frac{4\pi R^2}{\dot M_{\rm{Edd}}} \dot m_z \frac{dR}{R}\\
&=&\int^{r_{\rm{c}}}_{r_{\rm{i}}} \dot m_{\rm D}^*(\sqrt{C}-1)\frac{dr}{r}
\end {eqnarray}
with
\begin{equation}\label{mD}
\dot m_{\rm D}^*= \frac{4\pi R^2 \dot m_{\rm z}^*}{{\dot
M_{\rm{Edd}}}}=8.31\times 10^{-3}\alpha^{-7/5}\dot m^{7/5}r^{3/5}.
\end{equation}
The integrand $\dot m_{\rm D}^*(\sqrt{C}-1)$ gives the
contribution of logarithmic intervals in r. Its distribution
is shown in Fig. \ref{recond} for various values
of $\dot m$. The ordinate is the integrand relative to
the ADAF mass flow rate $\dot m$. The area underneath
each graph gives the relevant total condensation rate.

\begin{figure}
\centering
\includegraphics[width=6.cm]{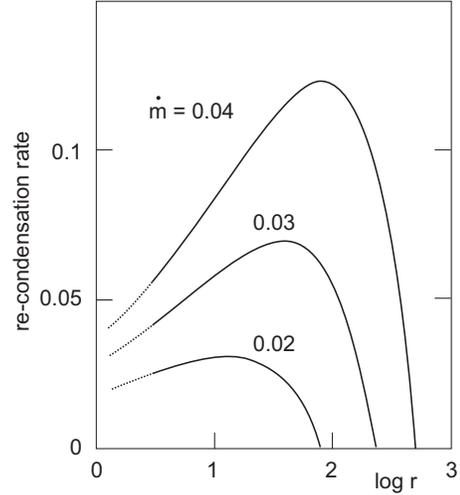}
\caption[.]{
{\label{recond}}
Re-condensation of gas from the ADAF onto
an inner disk as function of the distance $r$ (in units of $R_{\rm
S}$) for various mass flow rates $\dot m$
in the ADAF (in units of $\dot M_{\rm{Edd}}$). The curves
display the spatial distribution of contributions of
logarithmic intervals in $r$ as areas below the
curve. The ordinate is re-condensation rate relative to
the mass flow rate in the ADAF. Re-condensation occurs
at a larger distances for higher $\dot m$. Below $3 R_{\rm S}$ the
curves are dotted.}
\end{figure}

Introducing  $x=r/r_{\rm c}$ instead of $r$ in Eq.(\ref{condrate}) and 
making use of Eqs. (\ref{ccrit}), (\ref{rc}), and (\ref{mD}) one can 
evaluate the integral and obtain the total re-condensation rate as

\begin{equation}\label{totcondrate}
\frac{\mid \dot m_{\rm{cond}}\mid}{\dot m} 
= 2.34\times 10^{-2} \alpha^{-7} \dot m^2 
f(R_{\rm i}/R_{\rm c})
\end{equation}
with
\begin{equation}\label{fx}
f(x)=1-6x^{1/2} + 5x^{3/5}\,.
\end{equation}
The function $f$ approaches 1 when re-condensation begins at a 
radius $R_{\rm c}$ that is large compared to the inner cut-off 
radius $R_{\rm i}$. As the ADAF mass flow rate decreases 
$R_{\rm c}$ moves inward, $f$ and the re-condensation rate, 
i.e. the mass flow rate in the inner disk 
decrease until at a certain lower ADAF mass flow rate the outer disk edge
$R_{\rm c}$ reaches $R_{\rm i}$ where 
$f$ becomes 0 and the disk size has shrunk to zero. For even 
lower mass flow rates there is no 
re-condensation and the inner disk disappears. This intriguing 
behavior appears able to explain why Cyg X-1 does not 
show the hysteresis between hard/soft and soft/hard spectral 
state transitions that all transient X-ray sources display. 
We discuss this in the next section.

\subsection{Complete re-condensation?}

The model results depend on the parameters used to describe the 
ADAF. The dependence on the parameters $\alpha$ and $\dot m$ 
is particularly  strong for the re-condensation rate 
(Eq.(\ref{totcondrate})) and the radius at which re-condensation 
begins, i.e. the outer radius of an inner disk below the ADAF 
(Eq.(\ref{rc})). In the example given above both values 
are rather small. However, a small decrease of $\alpha$ and 
a slight increase of $\dot m$ can bring these values to 100\% 
re-condensation with re-condensation beginning already at 1000 
Schwarzschild radii (this is an extrapolation as our integral 
Eq.(\ref{condrate}) neglects changes of $\dot m$ by re-condensation).

This is relevant for standard disk evaporation 
models (e.g. Meyer et al. 2000). They predict that at distances 
of a few hundred to a thousand Schwarzschild radii there is a 
significant mass flow rate in an accretion disk corona, of a few 
percent of the Eddington value. If the mass accretion rate 
drops below that value, all the mass evaporates into the 
corona and from there on flows inward as a pure 
coronal flow, i.e. an ADAF. This yields a standard model for 
spectral state transitions. If the mass flow rate in the disk 
exceeds that critical value the disk is not truncated and 
continues to the last stable orbit. This accretion 
mode gives the characteristic  soft state spectrum of a 
multi-temperature black body. But even in this case 
the rather strong coronal flow at the distances of a few 
hundred Schwarzschild radii must exist. Why does it not 
flow on as a 2-temperature ADAF above the disks and shows as a
significant hard contribution to the soft state spectra? The answer 
from our intermediate state analysis 
is that re-condensation becomes complete for mass 
accretion rates as high as the coronal evaporation rate 
at a few hundred Schwarzschild radii. This explanation 
connects the processes of disk truncation and re-condensation. The 
upper part of  Fig. 1 illustrates the  situation.

\subsection {Hysteresis}
Even for lower mass accretion rates, when the outer 
disk already is truncated, re-condensation onto 
the remaining inner disk may be close to complete and still 
yield an apparent soft state in the innermost region. This 
extends the soft state to a lower mass accretion rate 
on the declining part of an outburst and leads to lower luminosity 
for the transition from the soft to the hard spectrum  in decline than for 
the opposite transition during the rise, i.e. hysteresis. Such 
hysteresis would be independent of, and add to the irradiation caused
hysteresis referred to in Sect. 9.

\section{Cyg X-1, a black hole accretor without hysteresis}

Cyg X-1 shows only moderate changes in the mass accretion rate,
attributed to the fact that  it is a wind accretor. This is very different 
from transient X-ray sources where the accretion rate increases and
decreases 100 fold during an outburst. Cyg X-1 stays in the hard 
state most of the time, but several spectral transitions to the soft  
state were observed. Because of the narrow range in mass accretion 
rate the hard state of Cyg X-1 is still intermediate in nature as the 
observations of Ibragimov et al. 
(2005) show.

Also by another fact Cyg X-1 differs from the transient sources. 
There is no noticeable hysteresis in Cyg X-1 (Zdziarski et al. 2002,
Zdziarski \& Gierli\'nski 2005). This was pointed out as a 
problem also by Lachowicz \& Czerny (2005).
Lightcurves of transient sources show a clear hysteresis in the transition 
luminosities (the soft/hard transition occurs at luminosities lower by
a factor of 3 to 5 than the hard/soft transition, e.g. in Aql X-1
(Maccarone \& Coppi 2003), GX 339-4 (Miyamoto et al. 1995, Nowak et
al. 2002, Zdziarski et al. 2004), XTE J1650-500 (Rossi et al. 2003),
XTE J1500-564 (Rodrigues et al. 2003). The different
luminosities at spectral transition can be understood as due to 
different maximal evaporation rates in the two states (triggering the 
transition) for either hard or soft irradiation of the corona by the 
central source (Meyer-Hofmeister et al. 2005). 

The new feature discussed in our work here, an inner re-condensation
disk, offers a  solution for the mystery of why Cyg X-1 has no
hysteresis while all the transients have it.

In transient sources the re-condensation disk recedes inward 
and finally disappears completely 
as the mass accretion rate continues to fall, as is the case
in the decline from outburst into quiescence. Then, 
no inner disk remains on which re-condensation could occur when 
the system moves into the next outburst. The spectrum remains 
hard  at increasing luminosity. Only when the accretion rate 
reaches the higher critical value for which the transition under hard 
irradiation occurs does it become soft.

Cyg X is different. When that system changes from a soft 
state excursion back to its usual hard state it does so with only a 
modest decrease of the mass accretion rate and never falls into 
deep quiescence. This allows an inner re-condensation disk to 
survive as the observations indicate. When now another episode 
of transition to a soft state occurs the mass accretion rate 
rises, re-condensation increases, the inner disk expands and 
carries an increasing disk mass flow until it smoothly merges 
with the outer disk to form the standard soft accretion 
state. Together with re-condensation the irradiation is completely 
reversible and no hysteresis occurs. This explanation relates 
the special character of the wind accretor to the otherwise enigmatic 
feature of "no hysteresis" in a natural way and seems a strong 
support for the re-condensation model.

\section{Discussion}
A more detailed numerical
investigation of the re-condensation process is desirable. The
present simple analytical model might serve to show the features and
their dependence on the parameters to be expected in this interesting
accretion mode.

\subsection{Interpretation of the intermediate state}

When the mass accretion rate gradually decreases further as 
a source declines from outburst the re-condensation 
rate onto the receding inner disk declines, a significant part 
of the accretion flow stays in the ADAF and shows  
characteristics of the hard state but modified by the 
existence of cool material in the form of a weakly accreting inner 
disk. From our analysis we find that the  
intermediate states during the outburst decline last longer for a slow 
decrease of the mass flow rate and less time for a fast decrease.

Observations for transient black hole X-ray sources show those
features. During the slow luminosity decline after the 1999 outburst of 
XTE J1859+226 (Corbel et al. 2004, Fig.9) the spectrum was
classified as due to an intermediate state. The steep luminosity
decline of GX 339-4 in 2003 (Remillard 2005, Fig.3) is
related to the very short intermediate state.

Short intermediate states of a few days 
might be caused by an inner disk disappearing in its viscous
time. Similar to these short states is the situation
during rise to outburst, where the inner edge of the disk moves inward
within a viscous time,  only an ADAF inside. If the mass flow rate in
the disk is always well below the maximal evaporation rate the disk is 
truncated all the time, and no thermal conduction affects the ADAF.
XTE J1118+480 is a well studied source which stays in the hard state
even during outburst. We would expect a disk truncation
at several hundred $R_{\rm S}$. But Esin et al. (2001) from 
multiwavelength observations  concluded that the disk truncation lies at about
55$R_{\rm S}$ in outburst. How can this be understood? The highest mass flow
rate during outburst was found to be about 0.02 $\dot M_{\rm{Edd}}$, the value
generally accepted for state transition (Maccarone 2003). This means
that 
the mass flow rate could have surpassed the critical rate for a short 
while and the outer disk began to extend inward. Re-condensation
immediately would begin. This re-condensation can then continue
even as the accretion rate drops below the critical rate, leading to
an intermediate state.

\subsection{Inner edge of the disk?}
The disk truncation in general varies with the mass accretion rate. For low
luminosity sources, X-ray binaries or AGN in a hard spectral state, the inner
edge of the outer disk was determined by fitting multiwavelength spectra, 
understood as caused by an ADAF and an outer disk. For
very low accretion rates large truncation radii were found, of the
order of 
$10^3$ to $10^4R_{\rm S}$ (e.g. Esin et al. 1997, Narayan et al. 1997, 
Narayan 2005). For X-ray binaries with higher mass accretion rates 
in phases of rise to an outburst or decay, from a
multicolor disk model smaller values for the inner 
disk radius were deduced (compare Sobczak et al. 2000, see examples in the
review of McClintock \& Remillard 2006). In a recent detailed 
investigation of the spectrum of Cyg X-1 Ibragimov et al.(2005) found 
truncation radii around 50-100 $R_{\rm S}$. Note that the mass flow rate
was not much below the transition rate, i.e. a typical intermediate
state. 

Comparing with the predictions of the standard evaporation model we 
find good agreement for the low luminosity
sources. The model also is consistent with no disk truncation 
for high mass flow rates. 

For the intermediate state  of Cyg X-1, Ibragimov 
et al. (2005) analyzed spectral observations 
in terms  of Compton reflection, the index of the hard
power law spectrum, the width of the Fe K$\alpha$ line, 
and an additional soft excess of a few keV of unknown origin.
With the correlation between photon index and amplitude of 
reflection they inferred disk cut-off  distances of 50-100 
$R_{\rm S}$, varying with the accretion rate.
Esin et al. (2001) inferred a similar truncation at 55 $R_{\rm S}$ for 
XTE J1118+480 in an apparently similar hard (-intermediate) 
state (see the discussion in the preceding subsection).

On the other hand, Miller et al. (2006a) argue for an untruncated 
disk extending to the last stable orbit during the low-hard state 
of the 2004 outburst of GX 339-4, at a luminosity of 5\% of the 
Eddington value. From a preliminary analysis of data 
from SWIFT J17335-0127 in the low-hard state of its 2005 outburst, 
Miller et al. (2006b) see indication for a similarly 
untruncated disk, here at a luminosity of one third of a per cent of the 
Eddington value. (Note that in optical thin emission the 
accretion luminosity can be lower than the 
corresponding accretion rate, both measured in their Eddington 
units.) This would suggest an untruncated 
re-condensation disk in these systems. The situation deserves further 
clarification.

The re-condensation model discussed here leaves open the 
question of if and where a remaining inner disk may itself be 
truncated. 
Neglected terms in the analysis, different assumptions about 
$\gamma$, and a more detailed calculation all can affect the 
outcome of the re-condensation rate estimates. One also 
cannot exclude that a parameter used to describe the ADAF 
varies with distance. The ratio of magnetic pressure to gas 
pressure, for example, could increase at smaller radii as 
accretion carries magnetic flux inwards. This decreases 
$\beta$ and $C$ (Eq.(\ref{C})). If C falls below one 
(Eq.(\ref{mdotzmdotz})), condensation turns into evaporation 
and the disk can become truncated before the final last stable 
orbit is reached. The effect of disk magnetic fields on evaporation in
the standard model is discussed in Meyer \& Meyer-Hofmeister (2002).

\subsection{Intermediate spectral states in disks around supermassive black
holes?}

The solutions discussed in this paper are independent of the mass $m$
of the black hole. Thus they should be applicable also to accretion on
supermassive black holes in AGN. Recently Jester (2005) studied
the distribution of AGN bolometric luminosities and black hole masses
for objects from the SDSS spectroscopic quasar survey to test the existence
of two different accretion modes (as predicted theoretically, Narayan et
al. 1998) and found a change of mode at an  Eddington-scaled accretion
rate of about 0.01. Markowitz and Uttley (2005), comparing power 
density functions pointed out the analogy of low luminosity AGN to 
low/hard state black hole X-ray binaries. This suggests that some of
the AGN accretors should be found in an intermediate state like their
galactic stellar mass counterparts. To actually observe a transition
through the different accretion states might be difficult as dynamical
and viscous timescales scale with the mass of the black hole.
Yuan et al. (2004) compared ROSAT and XMM Newton observations of 386
sources and found in one of them, the Seyfert-LINER galaxy NGC 7589, 
an increase in X-ray flux by a factor of $>$10 over a time of 5
years, a timescale of the order of the diffusion time at a  
distance of a few hundred Schwarzschild radii from a $10^7 M_\odot$
black hole. As the XMM high state luminosity was estimated to be a 
few percent of the Eddington value the authors suggest that the system
might have been found during a spectral transition.

\section{Conclusions}

We have presented a simple analytical model for re-condensation of gas from 
an ADAF onto a disk below under the action of thermal conduction and 
radiation. The model is able to explain how cool gas can exist in the close 
neighborhood of an accreting black hole, after a standard accretion disk 
has become truncated and an ADAF has formed,
a situation that occurs when the mass flow rate decreases during the 
transition from soft to hard spectral state of galactic black hole X-ray 
binaries. A fairly complete re-condensation of the ADAF into the disk 
would extend an inner thin disk accretion with its soft spectral state 
to lower mass accretion rates, and suggests a  hysteresis effect 
independent of and in addition to the irradiation effect 
discussed earlier. The model also clarifies why in soft state 
the considerable coronal flow at distances of several hundred 
Schwarzschild radii predicted theoretically does not continue as a hot 
2-temperature flow to the interior. Such a flow would appear as a hard 
contribution in the spectrum that is not observed. The model receives 
considerable support from being able to resolve the mystery of why 
Cyg X-1 does not show hysteresis between the luminosity at 
soft/hard and hard/soft spectral transition that all X-ray transient 
sources display. 

This simple model deserves further, more detailed investigation.
}

\begin{acknowledgements}
We thank Marat Gilfanov for helpful discussions. B.F. Liu
acknowledges support by
the National Natural Science Foundation of China (NSF-10533050)
and the BaiRenJiHua program of the Chinese Academy of Sciences.
\end{acknowledgements}

\end{document}